\documentclass[aps,prl,twocolumn,showpacs,preprintnumbers,nofootinbib,amsmath,amssymb]{revtex4-2}
\usepackage[pdftex]{graphicx}
\usepackage{latexsym,amsmath,amssymb,lmodern,float,url}
\usepackage{natbib}  
\usepackage[pdftex,bookmarks,linktocpage,pdfpagelabels,plainpages=false,hyperfigures,linkcolor=blue,citecolor=blue,urlcolor=blue]{hyperref} \usepackage{xspace}
\usepackage{courier}
\hypersetup{colorlinks=true}
\usepackage{bm}
\usepackage{dcolumn}
\usepackage{amsmath}
\usepackage{amssymb}
\usepackage{color}
\usepackage{xcolor}
\usepackage{soul}
\usepackage{url}
\usepackage{float}

\usepackage{aas_macros}
\usepackage[normalem]{ulem}
\usepackage{comment}
\usepackage{tikz-feynman}

\def\Eq#1{Eq.~(\ref{#1})}
\newcommand{\Eqs}[2]{Eqs.~(\ref{#1}) and (\ref{#2})}
\newcommand{\EqsMany}[2]{Eqs.~(\ref{#1} - \ref{#2})}
\def\Fig#1{Fig.~(\ref{#1})}

\def\H0{$H_0$}
\def\si8{$\sigma_8$}
\def\S8{$S_8$}

\newcommand{\Cite}[1]{Ref.~\cite{#1}}

\graphicspath{{./figs/}}

\begin{document}

\title{Mildly boosted dark matter annihilation and reconciling indirect galactic signals}

\author{Steven~J.~Clark} 
\email{sclark@hood.edu}
\affiliation{Hood College, Frederick, MD 21701, USA}

\begin{abstract}
The galactic center excess is a possible non-gravitational observation of dark matter; however, the canonical dark matter model (thermal freeze-out) is in conflict with other gamma-ray observations, in particular those made of the Milky Way's satellite dwarf galaxies. Here we consider the effects of a two-component dark matter model which results in minimally boosted particles that must remain bound to their host galaxy in order to produce an observational signal. This leads to a signal that is heavily dependent on galactic scale and can help reconcile the differences in the galactic center and dwarf galaxy measurements under the dark matter paradigm.
\end{abstract}

\maketitle 

\section{Introduction}
The galactic center excess (GCE) \cite{Goodenough:2009gk, Hooper:2010mq, Fermi-LAT:2017opo} is a flux of gamma rays originating from the center of the Milky Way galaxy that is higher than predictions from astrophysical processes. One possible interpretation is that it is due to dark matter (DM) interactions with Standard Model particles (SM); if correct, this interpretation would be the first non-gravitational detection of DM \cite{Calore:2014xka, Agrawal:2014oha, Calore:2014nla}. However, the DM parameter space that best correlates with the DM interpretation is also in conflict with other measurements; in particular, it is in conflict with similar gamma-ray measurements of the Milky Way's satellite spherical dwarf galaxies (dSph) \cite{Geringer-Sameth:2011wse, Fermi-LAT:2011vow, Fermi-LAT:2013sme, Geringer-Sameth:2014qqa, Fermi-LAT:2015att, DiMauro:2021qcf}.

If the GCE does originate from DM interactions, then reconciling these two observations can shed light on DM properties. Multiple models have attempted to address this difference in light of the DM proposal. Approaches frequently revolve around modifying the SM spectra through different particle productions \cite{Dutta:2015ysa, Cuoco:2017rxb} and adjusting the astrophysical interaction rates (commonly termed the J-factor). Some approaches for altering the J-factor include interactions with various velocity dependencies \cite{Boddy:2017vpe, Petac:2018gue, Boddy:2018ike, Boddy:2019wfg, Boddy:2019qak, Boucher:2021mii} or signals originating from secondary highly boosted DM \cite{Rothstein:2009pm, Finkbeiner:2014sja, Gori:2018lem, Agashe:2020luo}. For approaches that modify the J-factor, the central concept is that there is an inherent difference in the two environments (small and large galaxies) which leads to galactic dependencies not captured in the canonical value \cite{Agashe:2020luo}. In order to reconcile the GCE and dSph signals, this would require either an enhancement in larger galaxies or a suppression in smaller ones.

Expanded dark sectors offer a possible approach at addressing these signals by introducing dynamics that play a crucial role in DM distributions. Note that galactic DM is non-relativistic; this implies that processes that impart small increases to a particle's kinetic energy, compared with its rest mass, can lead to the particle achieving escape velocity, $v_{\rm esc}$, from the host galaxy. Because $v_{\rm esc}$ is dependent on galactic size, the requisite energy is lower in smaller galaxies and thus easier to escape. If this ``boosted'' DM particle is SM active, larger fractions of escaping boosted DM correspond with lower observational galactic signals. This leads to an overall suppression in the observed rates that is more pronounced in smaller galaxies, thus providing a mechanism to reconcile the GCE and dSph results.

In this work, we investigate galactic signal rates from multi-component DM models where the SM active components are created with a mild boost from a dominant SM inert portion. These types of models lead to a strong galactic dependence in observational rates with a rapid transition where galaxies above a critical scale experience minimal alterations to canonical rates while those below can experience strong to total suppression.

\section{Model and Bounded Fraction}
For illustrative purposes, we consider a basic two component dark matter toy model similar to \cite{Agashe:2020luo} consisting of $\chi_1$ and $\chi_2$ with mass relationships $m_1>m_2$ and $m_1/m_2 \approx 1$. $\chi_1$ annihilates to $\chi_2$ while $\chi_2$ annihilates to standard model particles.
\begin{align}
\chi_1 \chi_1 &\rightarrow \chi_2^{\mathrm b} \chi_2^{\mathrm b} \label{eq:chi1_ann} \\
\chi_2 \chi_2 &\rightarrow {\rm SM} \label{eq:chi2_ann}
\end{align}
where the ``b'' superscript indicates that the $\chi_2$s are produced with extra kinetic energy. We assume $\chi_1$ annihilation is weak allowing for $\chi_1$ to serve as the dark matter candidate while the $\chi_2$ annihilation rate is comparatively much stronger. From the perspective of a galaxy, $\chi_1$s comprise the majority of the dark matter, and $\chi_2$s are produced through $\chi_1$ annihilation. These $\chi_2$ will either be produced with sufficient velocity to overcome the gravitational potential and escape the galaxy, or they will remain bound. We will assume that if they achieve escape velocity, the $\chi_2$ annihilation coupling is small enough that they escape without further interaction (for annihilation occurring while escaping the galaxy, see \Cite{Agashe:2020luo}). If $\chi_2$s do not achieve escape velocity, they persist in the host galaxy until they annihilate with another $\chi_2$. In this setup, the galactic $\chi_2$ population fluctuates until it reaches a steady state solution, balancing between $\chi_2$ injections from the first interaction (\Eq{eq:chi1_ann} adjusted by the fraction that escape the galaxy) and depletion from the second annihilation (\Eq{eq:chi2_ann}) producing a SM signal similar to canonical DM annihilation.\footnote{Because $\chi_2$ will only experience a mild boost, their annihilation spectra will be identical to similar final products as canonical DM annihilation.} For this work, we assume that cross-sections are sufficient for all galaxies to reach this equilibrium state. The observable signal is directly proportional to the rate of $\chi_2$ annihilation, and when at equilibrium, it is also proportional to the $\chi_2$ injection rate.

\begin{figure}[h!]
\begin{tikzpicture}
\begin{feynman}
 
\vertex (b1) at (0,0);
\vertex (a1) at (-2,0);
\vertex (a2) at (2,0);
\vertex (a3) at (0.8944,1.7889);
\vertex (a4) at (-0.8944,-1.7889);
\vertex (c1) at (0,-2);
\vertex (c2) at (0,2);
\node at (1.4*1/8,1.4*2/4){$\theta$};
\node at (1.5,0.25){$\chi_1\colon\,v_1$};
\node at (-1.5,-0.25){$\chi_1\colon\,v_1$};
\node at (1,2){$\chi_2\colon\,v_2$};
\node at (-1,-2){$\chi_2\colon\,v_2$};

\diagram* {
(a1) -- [fermion,thick] (b1),
(a2) -- [fermion,thick] (b1),
(b1) -- [fermion,thick] (a3),
(b1) -- [fermion,thick] (a4),
(c1) -- [scalar] (c2)
};

\draw[->] (2.2cm,-1cm) -- (2.2cm,1cm);
\node at (2.5cm, 0){$v_c$};

\end{feynman}
\end{tikzpicture}
\caption{$\chi_1$ annihilation in the center of mass (COM) frame. Annihilation products that move in the same direction as the COM receive an increase to their velocity when converting to the galactic reference frame while those moving in the opposite direction are decreased. Depending on $v_2$, $v_c$, and the gravitational potential $\Phi$, a minimum $\theta$ is required to remain bound to the galaxy. If the boosting velocity is too large, all $\chi_2$ particles achieve $v_{\rm esc}$.}
\label{fig:cm}
\end{figure}
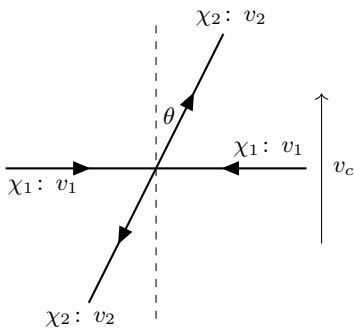

Two quantities are required to determine the rate of $\chi_2$ injection into the host galaxy: the base rate of $\chi_1$ annihilation producing $\chi_2$s and the $\chi_2^{\rm b}$ fraction from any particular $\chi_1$ annihilation that does not achieve escape velocity. We first discuss the fraction that does not reach $v_{\rm esc}$. \Fig{fig:cm} shows $\chi_1$ annihilation in the center of mass frame (COM). $v_{1,2}$ corresponds to the velocity of $\chi_{1,2}$ in the frame while $v_c$ is the velocity of the center of mass with respect to the galactic frame. $\theta$ is the angle between $v_c$ and $v_2$.

Using conservation of energy, it is easy to show that
\begin{align}
v_2^2 & = v_1^2 + \left(1-\frac{v_1^2}{c^2}\right)\left(1-\frac{m_2^2}{m_1^2}\right)c^2 \nonumber \\
& = v_1^2 + \Delta v^2 \label{eq:v_2^2}
\end{align}
where $\Delta v^2 \approx (1-m_2^2/m_1^2)\,c^2$ in the non-relativistic limit. In the COM frame, both $\chi_2$s have the same velocity; however, in the galactic reference frame, a difference develops depending on their orientation with $v_c$. In the galactic frame, denoted by subscript ``$g$'', the daughter particle velocity is
\begin{equation}
v_{2,g}^2 = v_2^2 + v_c^2 + 2 v_2 v_c \cos \theta
\label{eq:v_2^2_g}
\end{equation}
where we have assumed that all velocities are non-relativistic. It should also be noted that in the non-relativistic limit $v_1 = v_r/2$ where $v_r$ is the relative velocity between the two parent $\chi_1$ particles. This is true for both the COM and galactic reference frames.

For $\chi_2$ to remain bound to the galaxy, the total energy must be negative; this leads to the relationship $\Phi+v_{2,g}^2/2<0$ where $\Phi$ is the gravitational potential of the host galaxy. Combining this relationship with \Eqs{eq:v_2^2}{eq:v_2^2_g} as well as $v_1=v_r/2$, we arrive at the condition 
\begin{equation}
\cos \theta < \cos \theta_{\rm min} = -\frac{2 \Phi + v_c^2 + v_r^2/4 + \Delta v^2}{2 v_c \sqrt{v_r^2/4 + \Delta v^2}} \label{eq:theta_min}
\end{equation}
where $\theta_{\rm min}$ is the minimum angle that $v_2$ must make with $v_c$ in order for the $\chi_2$ to remain bound to the galaxy. If the right hand side of \Eq{eq:theta_min} is greater than 1, then the condition is always satisfied, all products are bound, and $\cos \theta_{\rm min} = 1$; if it is less than $-1$, $\cos \theta_{\rm min} = -1$ and all products escape. Note that \Eq{eq:theta_min} is valid only for $v_c>0$. For $v_c=0$, the binding condition is $-2\Phi>v_r^2/4 + \Delta v^2$; otherwise, all products escape.

For simplicity, we assume that $\chi_1$ annihilation is isotropic in the COM frame. In this isotropic example, the bound on $\cos \theta$ translates to the fraction of bounded annihilation products ($f_{\rm bound}$) through the fractional area of a 2-sphere between $\theta_{\rm min} \leq \theta \leq \pi$.
\begin{equation}
f_{\rm bound} =\frac{1}{2}\int_{\theta_{\rm min}}^\pi \sin \theta \,{\rm d}\theta = \frac{1+\cos \theta_{\rm min}}{2}
\end{equation}

\section{Initial Annihilation Rate}
To determine the initial annihilation rate from $\chi_1$, we follow the approach from \Cite{Boddy:2019wfg} with the addition of $f_{\rm bound}$ as discussed above to restrict the effective rate to include just the fraction of daughter particles which remain bound to the galaxy. For this work, the relevant quantity is $P_n^2(\hat{r})$. In canonical velocity independent annihilation rates, $P_n^2(\hat{r})$ is analogous to the square of the DM density ($\rho^2$) and evaluated from the DM velocity distribution, capturing the relative rate of annihilation occurring at a particular position in the galaxy. 
\begin{align}
P^2_n(\hat{r}) \equiv & \int d^3\hat{v}_1 d^3\hat{v}_2 |\hat{\bm{v}}_1-\hat{\bm{v}}_2|^n \nonumber \\
& \times \hat{f}(\hat{r},\hat{v}_1) \hat{f}(\hat{r},\hat{v}_2)f_{\rm bound}(\hat{\bm{v}}_1,\hat{\bm{v}}_2,\Delta \hat{v}) \nonumber \\
= & \; 8\pi^2 \int_0^\infty d\hat{v}_1 \int_0^\infty d\hat{v}_2 \int_{|\hat{v}_1-\hat{v}_2|}^{\hat{v}_1+\hat{v}_2} d\hat{v}_r \; \hat{v}_1 \hat{v}_2 \hat{v}_r^{n+1} \nonumber \\
& \times \hat{f}(\hat{r},\hat{v}_1) \hat{f}(\hat{r},\hat{v}_2) f_{\rm bound}(\hat{v}_1,\hat{v}_2,\hat{v}_r,\Delta\hat{v}) \label{eq:P^2}
\end{align}
Subscripts correspond to the two individual parent $\chi_1$ particles involved in the annihilation. (Note that this differs from the preceding section where subscripts indicated different particle species.) $\hat{f}(\hat{r},\hat{v})$ is the phase-space distribution for $\chi_1$ in the galaxy assumed here to be isotropic. $\hat{v}_r = |\hat{\bm{v}}_1-\hat{\bm{v}}_2|$ is the relative velocity between the two parent particles with $n$ being a model parameter. $f_{\rm bound}(\hat{v}_1,\hat{v}_2,\hat{v}_r,\Delta\hat{v}) = f_{\rm bound}(\hat{\bm{v}}_1,\hat{\bm{v}}_2,\Delta \hat{v})$ is the fraction of $\chi_2$ daughters bound to the host galaxy in an annihilation. Also note the relationship $4 \hat{v}_c +\hat{v}_r = 2(\hat{v}_1+\hat{v}_2)$.

For convenience in analyzing multiple galaxies later in this work, we have introduced the scaled distances, densities, and velocities in \Eq{eq:P^2}
\begin{equation}
\hat{r} \equiv \frac{r}{r_s}, \quad \hat{\rho} \equiv \frac{\rho}{\rho_s}, \quad {\rm and} \quad \hat{v} \equiv \frac{v}{\sqrt{4\pi G \rho_s r_s^2}}
\end{equation}
where $r_s$ and $\rho_s$ are the galactic scale and density parameters, and $G$ is the gravitational constant. Furthermore, the scaled phase-space distribution and gravitational potential are
\begin{align}
\hat{f}(\hat{r},\hat{v}) & = \left(4 \pi G \right)^{3/2} \rho^{1/2}r_s^3 f(r,v) \\
\hat{\Phi} & = \frac{\Phi}{4\pi G \rho_s r_s^2}
\end{align}
where $\hat{\rho}(\hat{r}) = \int d^3v \hat{f}(\hat{r},\hat{v})$.

As stated before, \Eq{eq:P^2} encapsulates the relative rate of DM annihilation, and $n$ captures the velocity dependence. For this work, we consider only the velocity independent interaction $n=0$ and leave $n\neq 0$ for future studies. For $n=0$ and $f_{\rm bound}=1$, \Eq{eq:P^2} reduces to $\hat{\rho}^2$ as expected.

J-factors (a measure of the expected flux) adjust the annihilation rate by accounting for the distance to the object and the observation window through a line of sight (l.o.s.) and region of interest (ROI) integration over $P_n^2$.
\begin{equation}
{\text{J-factor}} = \rho_s^2 \int_{\rm l.o.s.} d\ell \int_{\rm ROI} d\Omega \; P_n^2 (r/r_s)
\end{equation}

\subsection{Potentials and Phase-Space Distributions}
From a known DM distribution, the potential can be calculated using Newtonian gravity. If the distribution is spherically symmetric, then the scaled potential is \cite{Widrow:2000dm, Strigari:2010un, Boddy:2019wfg}
\begin{equation}
\hat{\Phi}(\hat{r}) = -\int_{\hat{r}}^\infty \frac{dx}{x^2} \int_0^x dy y^2 \hat{\rho}(y)
\end{equation}

In addition, if the DM velocity distribution is assumed to be isotropic and the halo is in equilibrium, the velocity distribution can also be determined from the density and the potential in terms of $E$, the energy per unit mass through \cite{Boddy:2019wfg}
\begin{align}
\hat{f}(\hat{E}) & =\frac{1}{\sqrt{8}\pi^2} \int_{\hat{E}}^0 \frac{d^2\hat{\rho}}{d\hat{\Phi}^2}\frac{d\hat{\Phi}}{\hat{\Phi}-\hat{E}} \\
\hat{E}(\hat{r},\hat{v}) & =\frac{E}{4\pi G\rho_s r_s^2} = \frac{\hat{v}^2}{2}+ \hat{\Phi}(\hat{r})
\end{align}
where we assume $v$ and $E$ go to zero at $r=\infty$. In this manner, we are able to define a fully self-consistent velocity distribution for the DM halo. The density can be found from the velocity distribution through
\begin{align}
\hat{\rho}(\hat{r}) & = 4\pi \int_0^{\sqrt{-2\hat{\Phi}(\hat{r})}} d\hat{v} \hat{v}^2 \hat{f}(\hat{r},\hat{v}) \\
& = 4\sqrt{2} \pi \int_{\hat{\Phi}(\hat{r})}^0 d\hat{E} \hat{f}(\hat{E}) \sqrt{\hat{E}-\Phi(\hat{r})}
\end{align}
where $\hat{f}(\hat{E}) = \hat{f}(\hat{r},\hat{v})$. For this work, we assume the NFW profile $\hat{\rho}(\hat{r}) = \{\hat{r}(1+\hat{r}^2)\}^{-1}$ \cite{Navarro:1995iw} for all galaxies. This leads to the gravitation potential $\hat{\Phi}(\hat{r}) = - \ln{(1+\hat{r})/\hat{r}}$ where $\ln(x)$ is the natural logarithm. From $\hat{\rho}(\hat{r})$ and $\hat{\Phi}(\hat{r})$, we solve for $\hat{f}(\hat{r},\hat{v})$ numerically.

\section{Adjusted Annihilation Rates}
\begin{figure}[h!]
\centering
\includegraphics[width=0.98\columnwidth]{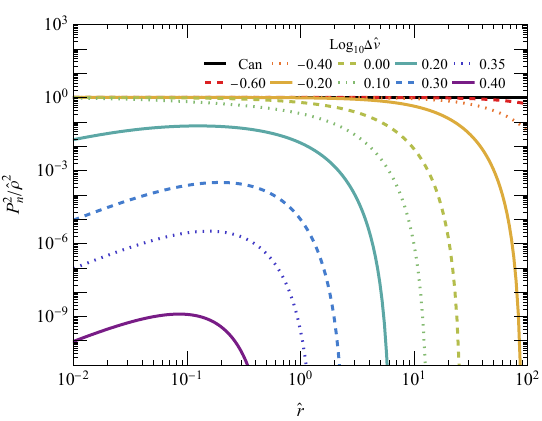}
\caption{Fraction of $\chi_2$ products from a $\chi_1$ annihilation that remain gravitationally bound to the host galaxy ($P_n^2/\hat{\rho}^2$) for various kick velocities, $\Delta\hat{v}$. ``Can'' refers to the canonical dark matter model with $P_n^2 = \hat{\rho}^2$ and is also equal to the total number of $\chi_1$ annihilation events in all models. As $\Delta\hat{v}$ increases, the level of suppression increases at all scales with larger radial distances more strongly affected. For $\Delta\hat{v} \gtrsim 1$, shorter distances also become heavily suppressed due to a lack of $v_c$ to allow adequate back-scattering. At $\Delta \hat{v} = \sqrt{8}$, all annihilation products escape the galaxy.} \label{fig:Adjusted_Rate}
\end{figure}

Due to the boost received during the first annihilation, only a fraction of $\chi_2$ remains bound to the galaxy. \Fig{fig:Adjusted_Rate} shows $P_n^2/\hat{\rho}^2$ for various kick velocities. The ratio between $P_n(\hat{r})^2$ and $\rho(\hat{r})^2$ is effectively the number of $\chi_2$ daughters that remain bound to the galaxy and the number that are produced. This ratio captures the fraction of the first annihilation products that are bound to the galaxy and which can participate in future interactions.

As would be expected from \EqsMany{eq:v_2^2}{eq:theta_min}, for $\Delta \hat{v} \ll 1$, all products remain in the host galaxy. This is due to the minimal changes to the energy distribution in the system. As $\Delta \hat{v}$ increases, escape occurs at large galactic radii due to the lower gravitational potential where even a small velocity change can provide the required energy. With increasing $\Delta \hat{v}$, the outer edges continue to become suppressed with the development of a critical radius above which there is total suppression.

This total suppression is due to the kick velocity providing the required energy for all valid velocity combinations to escape at the particular radius. From \Eq{eq:v_2^2_g}, $v_{c,{\rm max}}(r) = \Phi(r)$, and $\cos \theta = -1$, it is easy to find this maximum kick velocity where all annihilation products are unbound from the galaxy,
\begin{equation}
\Delta \hat{v}^2_{\rm max} < -8\hat{\Phi}.
\label{eq:Delta_v_max}
\end{equation}
For each curve shown in \Fig{fig:Adjusted_Rate}, this value is observed by the location of the sharp right cutoff.

This pattern continues until $\Delta \hat{v} \approx 1$ where increased suppression begins at small radii. This suppression is due to a lack of $v_c$ to allow for back-scattering events ($\cos \theta < 0$) that sufficiently reduce their energy. At $\Delta \hat{v} = \sqrt{2}$, the increase in kinetic energy is equal to the deepest portion of the NFW potential. This requires all bound products to be back-scattered with respect to $v_c$ in order to reduce their speed. The amount of energy reduction in back-scattering is more pronounced for larger $v_c$. At small galactic radii in the NFW distribution, a large proportion of the velocity distribution has low velocities compared with distributions at larger radii. This results in parent particles having lower average $v_c$ at small radii, leading to an incapability to sufficiently back-scatter to keep products bound to the galaxy even with the larger gravitational potential. Instead, these back-scattering events still have sufficient energy to escape. Rates at small radii thus experience a more dramatic suppression when compared to larger distances, and a peak in the rates is introduced for large $\Delta \hat{v}$ as observed in \Fig{fig:Adjusted_Rate}.

For $\Delta \hat{v}^2>-8\hat{\Phi}_{\rm max}$, all products escape and the galaxy is completely suppressed to 0. For the NFW profile, this kick velocity corresponds to $\Delta \hat{v}^2_{\rm max, \;NFW} = 8$ ($\log_{10} \Delta \hat{v}_{\rm max, \;NFW} \approx 0.45$).

\subsection{Total Rates}
\begin{figure}[h!]
\centering
\includegraphics[width=0.98\columnwidth]{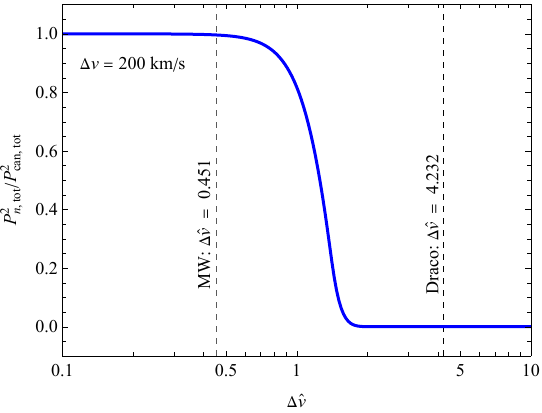}
\caption{Ratio between the total injection rate of bound $\chi_2$ daughter particles from $\chi_1$ annihilation ($P_{n, {\rm \; tot}}^2$) and the canonical annihilation rate. For $\Delta \hat{v} \ll 1$, the two are identical; however, there is a sharp drop at $\Delta \hat{v} \approx 1$ and the ratio reaches zero (no bound $\chi_2$ are injected) at $\Delta \hat{v} = \sqrt{8}$ for the NFW distribution. Also shown are $\Delta \hat{v}$ for the Milky Way (MW) and Draco galaxies for $\Delta v = 200 {\rm \; km/s}$. For this value of $\Delta v$, MW is close to the canonical result while Draco is deep in the totally suppressed regime. Other commonly studied dSph have $\Delta \hat{v}$ similar to Draco and would also be totally suppressed.} \label{fig:Adjusted_Rate_total}
\end{figure}

The total scaled annihilation rate for a galaxy can be found through $\int d^3 \hat{r} P_n^2(\hat{r}) = P_{n, {\rm \; tot}}^2$. In \Fig{fig:Adjusted_Rate_total}, we show the ratio between the galactic rate and the canonical result for the NFW distribution. Due to the assumption that $\chi_1$ and $\chi_2$ are in equilibrium, this ratio is equal to the change in expected signal from the second annihilation and the canonical result. For $\Delta\hat{v} \ll 1$, there is minimal variation from the canonical result as expected. At $\Delta\hat{v} \approx 0.5$, the rate begins to dramatically decrease such that it is negligible by $\Delta\hat{v} \approx 2$. At $\Delta\hat{v} = \sqrt{-8 \hat{\Phi}_{\rm max}}$, the rate reaches zero as no annihilation products are bound to the galaxy. As discussed earlier, for the NFW distribution, this occurs at $\Delta\hat{v} = \sqrt{8} \approx 2.83$.

For a single model, the kick velocity ($\Delta v$) will be constant for all galaxies; however, scaled velocities ($\Delta \hat{v}$) are galaxy dependent. Due to the sharp transition from full canonical expectations and a complete reduction to zero, some galaxies may experience almost no variation while others could experience dramatic departures from the canonical value. The main contributor to how different galaxies behave is their physical dimensions. In \Fig{fig:Adjusted_Rate_total}, we have included $\Delta \hat{v}$ for the Milky Way (0.451) and the dwarf galaxy Draco (4.232) assuming $\Delta v = 200 {\rm \; km/s}$.\footnote{For the Milky Way (Draco), we used $\rho_s = 0.345 \; (2.96) {\rm \; GeV/cm^3}$ and $r_s = 20 \; (0.728) {\rm \; kpc}$. \cite{Pace:2018tin} Other dSph galaxies commonly used in dark matter searches have similar values to Draco ($4 < \Delta \hat{v} < 6$).} For this choice of $\Delta v$, we would expect minimal alterations from the canonical signal for the Milky Way while expecting a complete suppression from Draco. Similar results will occur for other dSph galaxies. This suppression could account for the discrepancy between GC and Fermi dSph measurements.

Observing a signal from the Milky Way while experiencing a departure from the canonical result for dSph necessitates $\Delta v \sim 10 - 500 {\rm \; km/s}$, which corresponds to a mass splitting $\Delta m/m_1 \sim 10^{-9} - 10^{-6}$. This range is at the extreme of expected differences; a more conservative range would be $\Delta m/m_1 \sim 10^{-8} - 10^{-7}$ to allow for a sizable Milky Way signal while experiencing large variations from the canonical result for dSphs. Interestingly, this is the same mass splitting range as the 2cDM model needed to explain the missing galaxy, core-cusp, and too-big-to-fail problems of $N$-body simulations \cite{Todoroki:2017pdh, Todoroki:2020num, Chua:2020svq}; this is not completely unexpected due to the similarity in the models.

Care should be taken when interpreting these results, however, as they are a measure of the total annihilation rate of the galaxy. When converting to J-factors, they will be most accurate for distant galaxies, which the Milky Way is not. Further work is needed to understand how the secondary annihilation distribution in the galaxy will differ from canonical annihilation to identify if there are additional features when considering the angular J-factor along with the effect of galactic DM over-densities. In addition, more work is needed to better understand the equilibrium requirements necessary to reach the steady state situation assumed here and how it can differ in galactic environments.

Overall, this model introduces a mechanism that can explain the GCE as well as the lack of an observation in dSph under the DM interpretation. A characteristic feature to distinguish it from other models is a dramatic drop in excess SM products at a critical galactic scale. If measured, this scale can be used to measure the mass splitting between the two DM species.

\section*{Acknowledgements}
We thank Bhaskar Dutta and Savvas Koushiappas for helpful discussion and suggestions in the preparation of this work.

\bibliographystyle{utphys}
\bibliography{manuscript.bib}

\end{document}